\documentstyle{article}
\title{Euclidean scalar Green functions  near the black hole and black brane horizons }
\author{ Z. Haba\\Institute of Theoretical Physics, University of Wroclaw,
\\50-204 Wroclaw, Plac Maxa Borna 9, Poland\\e-mail:zhab@ift.uni.wroc.pl}
\date{PACS numbers 04.62+v,04.70.Dy }
\begin{document}
\maketitle
\begin{abstract}
We discuss  approximations of the Riemannian geometry near the
horizon. If a $D+1$ dimensional manifold ${\cal N}$ has a bifurcate
Killing horizon then we  approximate ${\cal N}$ by a product of the
two dimensional Rindler space ${\cal R}_{2}$ and a $D-1$ dimensional
Riemannian manifold ${\cal M}$. We obtain approximate formulas for
scalar Green functions. We study the behaviour of the Green
functions near the horizon and their dimensional reduction.
 We show that if ${\cal M}$ is compact
 then   the Green function near the horizon
can be approximated by
 the Green function of the two-dimensional quantum field theory.
 The correction term is exponentially small  away from the
 horizon. We extend the results to black brane solutions of
 supergravity in ten and eleven dimensions.
 The near horizon geometry can be approximated by ${\cal N}=AdS_{p}\times S_{q}$.
 We discuss the Euclidean Green functions on ${\cal N}$ and their behaviour near the horizon.
 \end{abstract}
\section{Introduction}
 The Hawking
radiation shows that quantum phenomena accompanying the motion of
a quantum particle in a neighborhood of a black hole require a
description in the framework of relativistic quantum theory of
many particle systems. It seems that
 quantum field theory supplies a proper method for a treatment of a varying number of particles.
 Quantum
field theory can be defined by means of Green functions.
 In the Minkowski space the locality and Poincare invariance
 determine the Green functions and allow a construction of free quantum fields.
  In the curved space  the Green function is not unique. The
non-uniqueness can be interpreted as a non-uniqueness of the
physical vacuum \cite{fulling}\cite{davis}. There is less ambiguity
in the definition of the Green function on the Riemannian manifolds
(instead of the physical pseudo-Riemannian ones).  The Euclidean
approach appeared successful when applied to the construction of
quantum fields on the Minkowski space-time \cite{jaffe0}. We hope
that such an approach will be fruitful in application to a curved
background as well. In contradistinction to the Minkowski space-time
an analytic continuation of Euclidean fields to quantum fields from
the Riemannian metric to the pseudoRiemannian one can be achieved
only if the manifold has an additional reflection symmetry
\cite{jaffe}(for a possible physical relevance of the reflection
symmetry see \cite{gibbons}).

The black hole can be defined in a coordinate independent way by
the event horizon. The event horizon is a global property of the
pseudoRiemannian manifold. It is not easy to see what is its
counterpart  after an analytic continuation to the Riemannian
manifold. Nevertheless, there is a proper substitute:the bifurcate
Killing horizon. As proved in \cite{racz} a manifold with the
Killing horizon  can be extended to the manifold with a bifurcate
Killing horizon. Moreover, there always exists an extension with
the wedge reflection symmetry \cite{kay} which seems crucial for
an analytic continuation between pseudoRiemannian and Riemannian
manifolds.
 The bifurcate Killing horizon is a local
property which can be treated in local coordinates \cite{wald2}. In
local static coordinates close to the bifurcate Killing horizon the
metric tensor $g$
 tends to zero at the horizon \cite{israel}. This property
is preserved after a continuation to the Riemannian metric. In
sec.2 we  approximate the Riemannian manifold ${\cal N}$ with the
bifurcate Killing horizon  as ${\cal N}={\cal R}_{2}\times {\cal
M}_{D-1}$, where ${\cal R}_{2}$ is the two-dimensional Rindler
space and ${\cal M}$ enters the definition of the bifurcate
Killing horizon as an intersection of past and future horizons.
There is the well-known example of the approximation in the form
of a product: the four-dimensional Schwarzschild solution can be
approximated near the horizon by a product of the Rindler space
and  the two-dimensional sphere . However, we do not restrict
ourselves to metrics which are solutions of Einstein gravity .

 The Euclidean quantum fields are defined by Green functions.
 For an approximate metric near the bifurcate Killing horizon we consider an
equation for the scalar Green functions. We expand the solution
into eigenfunctions of the Laplace-Beltrami operator on ${\cal
M}$. If ${\cal M}$ is compact without a boundary then the Laplace
-Beltrami operator has a discrete spectrum starting from $0$ (the
zero mode). We show in sec.3 that the higher modes are damped by a
tunneling mechanism. As a consequence, the position of the point
on the manifold ${\cal M}$ becomes irrelevant. The Green function
near the bifurcate Killing horizon can be well approximated by the
Green function of the two-dimensional  free field. The splitting
of the Green function near the horizon into a product of the two
dimensional function and a function on ${\cal M}$ has been
predicted by Padmanabhan \cite{padma}. However, we obtain its
exact form. In sec.4 we discuss an application of our earlier
results \cite{haba-cqg} on the dimensional reduction of the Green
functions. In sec.5 we apply the method to black brane solutions
of string theory \cite{brane}\cite{duff} in 10 and 11 dimensional
supergravity which at the horizon have the geometry of
$AdS_{p}\times S_{q}$. We show that Euclidean quantum field theory
on the brane can be approximated by that on the hyperbolic space (
the Euclidean version of AdS). The propagator on the background
manifold of the black brane can be applied for a construction of
supergravity with the black brane as the vacuum state.
Anti-de-Sitter space is the homogeneous space of the conformal
group. Hence, at the level of the two-point functions we derive
the relation between supergravity and conformal field theory on
the boundary of of $AdS_{p}$ (which is $S_{p-1}$)\cite{witten}.

\section{An approximation at the bifurcate Killing horizon}We consider a $D+1$ dimensional Riemannian
manifold ${\cal N}$ with a metric $\sigma_{AB}$ and a bifurcate
Killing horizon. This notion assumes a symmetry generated by the
Killing vector $\xi$. Then, it is assumed that the Killing vector
is orthogonal to a (past oriented) $D$ dimensional hypersurface
${\cal H}_{A}$ and a (future oriented) hypersurface ${\cal H}_{B}$
\cite{wald2}. The Killing vector $\xi^{A}$ is vanishing
(i.e.,$\xi^{A}\xi_{A}=0$) on an intersection of ${\cal H}_{A}$ and
${\cal H}_{B}$ defining a $D-1$ dimensional surface ${\cal M}$
(which can be described as the level surface $f=const$) .  The
bifurcate Killing horizon implies that the space-time has locally
a structure of an accelerated frame, i.e., the structure of the
Rindler space \cite{rindler}. In \cite{racz} it is proved that the
space-time with a Killing horizon can be extended to a space-time
with the bifurcate Killing horizon. Padmanabhan \cite{padma}
describes such a bifurcate Killing horizon as a transformation
from a local Lorentz frame to the local accelerated (Rindler)
frame. In \cite{racz} it is  shown that the extension can be
chosen in such a way that the "wedge reflection symmetry"
\cite{kay} is satisfied.In the local Rindler coordinates  the
reflection symmetry is $X=(x_{0},y,z)\rightarrow
\tilde{X}=(x_{0},-y,z)$. The symmetry means that the metric
$\sigma_{AB}$ splits into a block form (we denote coordinates on
${\cal N}$ and  its indices by capital letters)
\begin{displaymath}
ds^{2}=
\sigma_{AB}dX^{A}dX^{B}=g_{00}dx^{0}dx^{0}+g_{11}dydy+2g_{01}dx^{0}dy+\sum_{j,k>1}g_{jk}dz^{j}dz^{k}
\end{displaymath}
 The bifurcate Killing horizon
 distinguishes
a two-dimensional subspace of the tangent space.  At the bifurcate
Killing horizon the two-dimensional metric tensor $g_{ab}$ is
degenerate. In the adapted coordinates such that
$\xi=\partial_{0}$ we have $g_{10}=0$ and the metric does not
depend on $x_{0}$. Then, $\det[g_{ab}]\rightarrow 0$ at the
horizon means that $g_{00}(y=0,z)=0$ or $g_{11}(y=0,z)=0$ . We
assume $g_{00}(y=0,z)=0$. As $g_{00}$ is non-negative its Taylor
expansion must start with $y^{2}$. Hence, if we neglect the
dependence of the two-dimensional metric $g_{ab}$ on $x_{0}$ and
on $z$ then we can write it in the form
\begin{equation} \begin{array}{l}ds_{g}^{2}=-y^{2}(dx^{0})^{2}+dy^{2}+\sum_{j,k\geq
2}g_{jk}(y,z)dz^{j}dz^{k}\cr \equiv y^{2}
\Big(-(dx^{0})^{2}+y^{-2}(dy^{2}+ds_{D-1}^{2})\Big)
\end{array}\end{equation}
If we neglect the dependence of $g_{jk}$ on $y$  near the horizon
then the metric $ds_{D-1}^{2}$ ( denoted $ds_{M}^{2}$ ) can be
considered as a metric
 on the $D-1$ dimensional  surface ${\cal M}$  being the common part of
 ${\cal H}_{A}$ and ${\cal H}_{B}$.  Hence, in eq.(1)
 ${\cal N}={\cal R}_{2}\times {\cal M}$ where ${\cal R}_{2}$
is the two-dimensional Rindler space. As an example of the
approximation of the geometry of ${\cal N}$
 we could consider the four dimensional Schwarzschild black hole when  ${\cal N}\simeq
 {\cal R}_{2}\times
 S_{2}$ ( quantum theory with such an approximation is discussed in \cite{bykowski})  .

We shall work with  Euclidean  version of the metric (1)
\begin{equation} \begin{array}{l}ds^{2}
=y^{2}(dx^{0})^{2}+dy^{2}+\sum_{j,k\geq 2}g_{jk}(0,z)dz^{j}dz^{k}
\end{array}\end{equation}

We consider the Laplace-Beltrami operator
\begin{displaymath}
\triangle_{N}=\frac{1}{\sqrt{\sigma}}\partial_{A}\sigma^{AB}\sqrt{\sigma}\partial_{B}
\end{displaymath}
on ${\cal N}$ ($\sigma=\det(\sigma_{AB})$).

In the approximation (2) we have (if $g_{jk}$ is independent of $y$)
\begin{equation}
 \triangle_{N}=y^{-2}\partial_{0}^{2}+y^{-1}\partial_{y}y\partial_{y}+\triangle_{M}
 =\triangle_{R}+\triangle_{M}\end{equation}
where $\triangle_{R}$ is the Laplace-Beltrami operator on the
two-dimensional Rindler space and $\triangle_{M}$ is the
Laplace-Beltrami operator for the metric
\begin{displaymath}
ds_{M}^{2}=\sum_{jk}g_{jk}(0,z)dz^{j}dz^{k}
\end{displaymath}We are interested in the calculation of the Green
functions
\begin{equation}(-\triangle_{N} +m^{2}){\cal G}_{N}^{m}=\frac{1}{\sqrt{\sigma}}\delta
\end{equation}
Then, eq.(4) for ${\cal N}={\cal R}_{2}\times{\cal M}$ reads
\begin{equation}
-(\partial_{0}^{2}+y\partial_{y}y\partial_{y}+y^{2}\triangle_{M}-y^{2}m^{2}){\cal
G}_{N}^{m}=y\frac{1}{\sqrt{g}_{M}}\delta
\end{equation}
 After  an exponential change of coordinates
\begin{equation}
y=\exp x_{1}
\end{equation}
eq.(5) takes the form
\begin{equation}
\Big(-\partial_{0}^{2}-\partial_{1}^{2}-\exp(2x_{1})(\triangle_{M}-m^{2})\Big){\cal
G}^{m}_{N}=g_{M}^{-\frac{1}{2}}\delta(x_{0}-x_{0}^{\prime})\delta(x_{1}-x_{1}^{\prime})\delta(z-z^{\prime})
\end{equation}
If ${\cal M}$ is approximated by $R^{D-1}$ then the metric (1) is
conformally related to the hyperbolic metric. This relation has its
impact on the form of the Green functions (5) as will be seen in
secs.3 and 4.

\section{Green functions near the bifurcate Killing horizon}
We investigate in this section the Green function (5) in $D+1$
dimensions under the assumption that ${\cal M}$ is $D-1$
dimensional compact manifold without a boundary. We introduce the
complete basis of eigenfunctions in the space
$L^{2}(dx_{0}dx_{1})$ of the remaining two  coordinates
\begin{equation}
(-\partial_{0}^{2}-\partial_{1}^{2}+\omega_{k}^{2}\exp(2x_{1})\Big)\phi_{k}^{E}(x_{0},x_{1})=E
\phi_{k}^{E}(x_{0},x_{1})
\end{equation}
where
\begin{equation}
\omega_{k}^{2}=\epsilon_{k}+m^{2}
\end{equation}
In eq.(8) $E$ denotes the set of all the parameters the solution
$\phi^{E}$ depends on. The solutions $\phi$ satisfy the
completeness relation
\begin{equation}
\int
d\nu(E)\overline{\phi}_{k}^{E}(x_{0},x_{1})\phi_{k}^{E}(x_{0}^{\prime},x_{1}^{\prime})
=\delta(x_{0}-x_{0}^{\prime })\delta(x_{1}-x_{1}^{\prime })
\end{equation}
with a certain measure $\nu$ and the orthogonality relation
\begin{equation}\int
dx_{0}dx_{1}\overline{\phi}_{k}^{E}(x_{0},x_{1})\phi_{k}^{E^{\prime}}(x_{0},x_{1})=
\delta(E-E^{\prime})\end{equation} where again the $\delta$
function concerns all parameters characterizing the solution.

If ${\cal M}$ is a compact manifold without a boundary then
$-\triangle_{M}$ has a complete discrete set of orthonormal
eigenfunctions \cite{chavel}
\begin{equation} -\triangle_{M}u_{k}=\epsilon_{k}u_{k}
\end{equation}
satisfying the completeness relation
\begin{displaymath}
\sum_{k}\overline{u}_{k}(z)u_{k}(z^{\prime})=g^{-\frac{1}{2}}\delta(z-z^{\prime})
\end{displaymath}
Solutions of Eq.(12) always have a zero mode $u_{k}=1$ (we normalize
the Riemannian volume element of ${\cal M}$ to $1$) corresponding to
$\epsilon_{k}=0$. We expand ${\cal G}$ (we drop the index $N$ in
${\cal G}$ because there is only one Green function in this section)
distinguishing the contribution of the zero mode
\begin{equation}\begin{array}{l}
{\cal G}^{m}(x_{0},x_{1},z;x_{0}^{\prime},x_{1}^{\prime},z^{\prime})
-g_{0}^{m}(x_{0},x_{1};x_{0}^{\prime},x_{1}^{\prime})=\sum_{k\neq 0}
g_{k}^{m}(x_{0},x_{1};x_{0}^{\prime},x_{1}^{\prime})
\overline{u}_{k}(z)u_{k}(z^{\prime})\end{array}\end{equation}where
\begin{equation}
g_{k}^{m}(x_{0},x_{1};x_{0}^{\prime},x_{1}^{\prime})=\int
d\nu(E)E^{-1}\overline{\phi}_{k}^{E}(x_{0},x_{1})\phi_{k}^{E}(x_{0}^{\prime},x_{1}^{\prime})
\end{equation}
$g_{k}^{m}$ is the kernel of the inverse of the operator
\begin{displaymath}
H=-\partial_{0}^{2}-\partial_{1}^{2}+\omega_{k}^{2}\exp(2x_{1})
\end{displaymath}
i.e.,
\begin{equation}
Hg_{k}^{m}(x_{0},x_{1};x_{0}^{\prime},x_{1}^{\prime})=\delta(x_{0}-x_{0}^{\prime})\delta(x_{1}-x_{1}^{\prime})
\end{equation} We have moved
the zero mode in eq.(13) from the rhs to the lhs. $g^{m}_{0}$ is
the solution of the equation
 \begin{equation}
(-\partial_{0}^{2}-\partial_{1}^{2}+m^{2}\exp(2x_{1}))g^{m}_{0}=\delta(x_{0}-x_{0}^{\prime})\delta(x_{1}-x_{1}^{\prime})
\end{equation}
 Taking the Fourier transform in $x_{0}$ we
express eq.(16) in the form
\begin{equation}
\tilde{H}\tilde{g}^{m}_{k}(p_{0},x_{1};,x_{1}^{\prime})=\delta(x_{1}-x_{1}^{\prime})
\end{equation}
where
\begin{equation}
\tilde{H}=p_{0}^{2}-\partial_{1}^{2}+\omega_{k}^{2}\exp(2x_{1})
\end{equation}
In order to obtain an approximate estimate on the behaviour of
$\tilde{g}_{k}^{m}$ we write $\tilde{g}_{k}^{m}$ in the  WKB form
\begin{equation}
\tilde{g}_{k}=\exp(-\omega_{k}W)
\end{equation}
Then, from eq.(15) for large $x_{1}$
\begin{equation}
\omega_{k}^{2}(\partial_{1}W)^{2}=\omega_{k}^{2}\exp(2x_{1})+p_{0}^{2}
\end{equation}
Hence, if $\omega_{k}\neq 0$ then $W\simeq \exp(x_{1})$ for large
$x_{1}$ showing that ${\cal G}^{m}-g_{0}^{m}$ is decaying
exponentially fast away from the horizon (if $m=0$ then
$\omega_{k}>0$ if $\epsilon_{k}>0$).
 We study this phenomenon in more detail now. First,
we write the solution of eq.(8) in the form
\begin{equation}
\phi_{k}^{E}=\exp(ip_{0}x_{0})\phi_{k}^{p_{1}}(x_{1})
\end{equation}where

\begin{equation}
(-\partial_{1}^{2}+\omega_{k}^{2}\exp(2x_{1}))\phi_{k}^{p_{1}}=p_{1}^{2}\phi_{k}^{p_{1}}
\end{equation}Now, $E=p_{0}^{2}+p_{1}^{2}$ and $d\nu=dp_{0}dp_{1}$ in
eqs.(10)-(11). When $\omega_{k}=0$ then the solution of eq.(22) is
the plane wave
\begin{displaymath}
\phi_{0}^{p_{1}}=(2\pi)^{-\frac{1}{2}}\exp(ip_{1}x_{1})
\end{displaymath}

 The normalized
solution of eq.(22, which behaves like a plane wave with momentum
$p_{1}$ for $x_{1}\rightarrow -\infty$ and decays exponentially
for $x_{1}\rightarrow +\infty$, reads
\begin{equation}
\phi_{k}^{p_{1}}=N_{p_{1}}K_{ip_{1}}(\omega_{k}\exp(x_{1}))
\end{equation}
where $K_{\nu}$ is the modified Bessel function of the third kind
of order $\nu$ \cite{grad}.

 This solution is inserted into the formulas (13)-(14)
for the Green function with the normalization (11)
\begin{equation}
\int_{-\infty}^{\infty}dx_{1}\overline{\phi}_{k}^{p_{1}}(x_{1})\phi_{k}^{p_{1}^{\prime}}(x_{1})
=\delta(p_{1}-p_{1}^{\prime})
\end{equation}
Hence (see \cite{fulling}\cite{sommer}\cite{sessa}),
\begin{equation} N_{p_{1}}^{2} =p_{1}\sinh(\pi
p_{1})\frac{2}{\pi^{2}}
\end{equation} Then,  performing the integral over $p_{0}$ in
eq.(14)
\begin{displaymath}
\int
dp_{0}\exp(ip_{0}(x_{0}-x_{0}^{\prime}))(p_{0}^{2}+p_{1}^{2})^{-1}=
\pi \vert p_{1}\vert^{-1}\exp(-\vert p_{1}\vert \vert
x_{0}-x_{0}^{\prime}\vert)
\end{displaymath}
we obtain
\begin{equation}\begin{array}{l}
{\cal
G}(x_{0},x_{1},z;x_{0}^{\prime},x_{1}^{\prime},z^{\prime})^{m}-
g_{0}^{m}(x_{0},x_{1};x_{0}^{\prime},x_{1}^{\prime})
=\frac{4}{\pi^{2}}\int_{0}^{\infty}dp_{1}\sinh(\pi
p_{1})\exp(-p_{1}\vert x_{0}-x_{0}^{\prime}\vert) \cr\sum_{k\neq
0}K_{ip_{1}}(\omega_{k}\exp(x_{1}))
K_{ip_{1}}(\omega_{k}\exp(x_{1}^{\prime}))
\overline{u}_{k}(z)u_{k}(z^{\prime})\end{array}\end{equation}

We expect that the rhs of eq.(26) is decaying fast to zero away
from the horizon. Each term on the rhs of eq.(26) is decaying
exponentially fast when $x_{1}\rightarrow +\infty$ because the
Bessel function $K_{\nu}$ is decreasing exponentially. We  wish to
show that the sum  is decreasing exponentially as well. This is
not a simple problem because we need some estimates on
eigenfunctions and eigenvalues of the Laplace-Beltrami operator on
${\cal M}$.

 Let us consider the simplest example ${\cal M}=S^{a}_{1}$  (the circle of radius $a$ which
can be related to the three-dimensional BTZ black hole
\cite{BTZ}). Then, $u_{k}(x_{2})=(2\pi
a)^{-\frac{1}{2}}\exp(\frac{i}{a}kx_{2})$ and
$\epsilon_{k}=a^{-2}k^{2}$ (here we denote the coordinate $z$ of
eq.(26) by $x_{2}$). We assume $m=0$ for simplicity of the
argument. Then, the sum over $k$ can be performed
 by means
of the representation of the Bessel function (if $m\neq 0$ then we
are unable to do the summation exactly but for our estimates this is
not necessary because only large eigenvalues are relevant for large
eigenvalues $\omega_{k}\simeq\epsilon_{k}$)
\begin{equation}
K_{i\nu}(u)=\int_{0}^{\infty}dt\exp(-u\cosh t)\cos(\nu t)
\end{equation}
We have
\begin{equation}
\begin{array}{l}\sum_{k=1}^{\infty}\exp\Big(-ku\cosh(t)-ku^{\prime}\cosh(t^{\prime})\Big)
\cos(\frac{k}{a}(x_{2}-x_{2}^{\prime}))=
\cr\Re\Big(\exp\Big(-u\cosh(t)-u^{\prime}\cosh(t^{\prime})+\frac{i}{a}(x_{2}-x_{2}^{\prime})\Big)\cr
\Big(1-\exp(-u\cosh(t)-u^{\prime}\cosh(t^{\prime})+\frac{i}{a}(x_{2}-x_{2}^{\prime}))\Big)^{-1}\Big)
\end{array}
\end{equation}
Inserting (28) in eq.(26)  and approximating the denominator in
eq.(28) by 1 we obtain an asymptotic estimate for large positive
$x_{1}$ and $x_{1}^{\prime}$ (large $y$ and $y^{\prime}$;this is
eq.(26) neglecting $\vert k\vert>1$)
\begin{equation}\begin{array}{l}
{\cal
G}^{0}(x_{0},x_{1},x_{2};x_{0}^{\prime},x_{1}^{\prime},x_{2}^{\prime})-
g_{0}^{0}(x_{0},x_{1};x_{0}^{\prime},x_{1}^{\prime})
\simeq\frac{8}{a\pi^{2}}\cos(\frac{x_{2}}{a}-\frac{x_{2}^{\prime}}{a})\cr
\int_{0}^{\infty} dp_{1}\sinh(\pi p_{1})\exp(-p_{1}\vert
x_{0}-x_{0}^{\prime}\vert)
K_{ip_{1}}(\frac{1}{a}\exp(x_{1}))K_{ip_{1}}(\frac{1}{a}\exp(x_{1}^{\prime}))\end{array}\end{equation}
We can  estimate the integral over $p_{1}$ if $\vert
x_{0}-x_{0}^{\prime}\vert>\pi$(if this condition is not satisfied
then the estimates are much more difficult because we must estimate
the behaviour of $K_{ip_{1}}(\exp(x_{1}))$ simultaneously for large
$p_{1}$ and large $x_{1}$). In such a case inserting the asymptotic
expansion of the Bessel function we obtain
\begin{equation}\begin{array}{l}
{\cal
G}^{0}(x_{0},x_{1},x_{2};x_{0}^{\prime},x_{1}^{\prime},x_{2}^{\prime})-
g_{0}^{0}(x_{0},x_{1};x_{0}^{\prime},x_{1}^{\prime})\cr
\simeq\frac{8}{a\pi^{2}}\sqrt{\frac{\pi}{2a}}\exp(-\frac{1}{2}x_{1}-\frac{1}{2}x_{1}^{\prime})\cr
\cos(\frac{x_{2}}{a}-\frac{x_{2}^{\prime}}{a})(\vert
x_{0}-x_{0}^{\prime}\vert-\pi)^{-1}
\exp\Big(-\frac{1}{a}\exp(x_{1})-\frac{1}{a}\exp(x_{1}^{\prime})\Big)\end{array}\end{equation}
where (for $m=0$) \begin{displaymath}
g_{0}^{0}(x_{0},x_{1};x_{0}^{\prime},x_{1}^{\prime})=-\frac{1}{4\pi}
\ln((x_{0}-x_{0}^{\prime})^{2}+(x_{1}-x_{1}^{\prime})^{2})
\end{displaymath}
is the solution of the equation
\begin{equation}
(-\partial_{0}^{2}-\partial_{1}^{2})g_{0}^{0}=
\delta(x_{0}-x_{0}^{\prime})\delta(x_{1}-x_{1}^{\prime})
\end{equation}
It follows that ${\cal G}^{0}$ close to the horizon tends
exponentially fast to the Green function for the two-dimensional
quantum field theory.

In a similar approach we treat the case ${\cal M}=S_{2}^{a}$,
where $S_{2}^{a}$ is the two-dimensional sphere with radius $a$.
This case is interesting because it describes the near horizon
geometry of the four-dimensional Schwarzschild black hole
\cite{becker}. The approximate near horizon geometry of the
four-dimensional Schwarzschild black hole is ${\cal N}={\cal
R}_{2}\times S_{2}^{a}$ (in $D+1$ dimensions this is ${\cal
N}={\cal R}_{2}\times S_{D-1}^{a}$) where ${\cal R}_{2}$ is the
two-dimensional Rindler space. Now, the formula (26) reads
\begin{equation}\begin{array}{l}
{\cal
G}_{N}^{m}(x_{0},x_{1},\theta,\phi;x_{0}^{\prime},x_{1}^{\prime},\theta^{\prime},\phi^{\prime})-
g_{0}^{m}(x_{0},x_{1};x_{0}^{\prime},x_{1}^{\prime})
=\frac{1}{\pi^{3}}\int_{0}^{\infty}dp_{1}\sinh(\pi
p_{1})\exp(-p_{1}\vert x_{0}-x_{0}^{\prime}\vert) \cr\sum_{l\neq
0}K_{ip_{1}}(\omega_{l}\exp(x_{1}))
K_{ip_{1}}(\omega_{l}\exp(x_{1}^{\prime}))
(2l+1)P_{l}(\cos\frac{\sigma}{a})\end{array}\end{equation} where
$\omega_{l}^{2}=l(l+1)a^{-2}+m^{2}$, $P_{l}$ is the Legendre
polynomial and $\sigma$ is the geodesic distance on $S_{2}$ . In the
spherical angles $(\theta,\phi)$\begin{equation}
(\cos\frac{\sigma}{a})(\theta,\phi;\theta^{\prime},\phi^{\prime})
=\cos\theta\cos\theta^{\prime}+\sin\theta\sin\theta^{\prime}\cos(\phi-\phi^{\prime})
\end{equation}
A finite number of terms on the rhs of eq.(32) is decaying
exponentially in $y=\exp x_{1}$ (because $K_{ip_{1}}$ is decaying
exponentially). Hence, it is sufficient to estimate the sum in
eq.(32) for $l\geq L$. For large $l$ we can use the approximation
$\omega_{l}\simeq \frac{l}{a}$. Applying the representation of the
Legendre polynomials
\begin{equation}
P_{l}(\cos\theta)=\frac{1}{\pi}\int_{0}^{\pi}(\cos\theta
+i\sin\theta\cos\phi)^{l}d\phi
\end{equation}
and the representation (27) of the Bessel functions we obtain
\begin{equation}\begin{array}{l}
\sum_{l\geq
L}\gamma^{l}(2l+1)\exp\Big(-\frac{l}{a}\exp(x_{1})\cosh(t)-\frac{l}{a}\exp(x_{1}^{\prime})\cosh(t^{\prime})\Big)\cr
=\Gamma^{L}\Big(2L+1-(2L-1)\Gamma \Big)
(1-\Gamma)^{-2}\end{array}\end{equation} where
\begin{displaymath}
\Gamma=\gamma\exp\Big(-\frac{1}{a}\exp(x_{1})\cosh(t)-\frac{1}{a}\exp(x_{1}^{\prime})\cosh(t^{\prime})\Big)
\end{displaymath}
and

\begin{equation}
\gamma=\cos\sigma+i\sin\frac{\sigma}{a}\cos\phi
\end{equation}
If in $(1-\Gamma)^{-2}$ we neglect $\Gamma$ (or expand it in a
power series of $\Gamma$) and apply the asymptotic expansion for
the Bessel functions then we can conclude that the sum starting
from $L$ is decaying as
$\exp(-\frac{L}{a}\exp(x_{1})-\frac{L}{a}\exp(x_{1}^{\prime}))$
for large positive $x_{1}$. Therefore, the behaviour of the rhs of
eq.(26) for large positive $x_{1}$ is determined by the lowest
non-zero eigenvalue. Taking only the term with the lowest non-zero
eigenvalue we obtain similarly as in eq.(30) (if $\vert
x_{0}-x_{0}^{\prime}\vert > \pi$ ) the approximation
\begin{equation}\begin{array}{l}
 {\cal
G}^{m}(x_{0},x_{1},\theta,\phi;x_{0}^{\prime},x_{1}^{\prime},\theta^{\prime},\phi^{\prime})-
g_{0}^{m}(x_{0},x_{1};x_{0}^{\prime},x_{1}^{\prime})\cr \simeq
\frac{12}{\pi^{3}}\sqrt{\frac{\pi}{2a}}
\exp(-\frac{1}{2}x_{1}-\frac{1}{2}x_{1}^{\prime})\cr(\vert
x_{0}-x_{0}^{\prime}\vert-\pi)^{-1} \exp\Big(-\sqrt{m^{2}+\frac{
2}{a^{2}}}(\exp(x_{1})+\exp(x_{1}^{\prime}))\Big)\cos\frac{\sigma}{a}\end{array}\end{equation}
where $g_{0}^{m}$ is the solution of eq.(16).

For general compact manifolds ${\cal M}$ we must apply some
approximations in order to estimate the infinite sums. We estimate
the rhs of eq.(13) for large $x_{1}$ and $x_{1}^{\prime}$ by means
of a simplified argument applicable when $z=z^{\prime}$ ,
$x_{1}=x_{1}^{\prime}$ and  $\vert u(z)\vert\leq C$. Then,
\begin{equation}\begin{array}{l}
\vert{\cal G}^{m}(x_{0},x_{1},z;x_{0}^{\prime},x_{1},z)-
g_{0}^{m}(x_{0},x_{1};x_{0}^{\prime},x_{1}) \vert\cr \leq
C^{2}\frac{4}{\pi^{2}}\int_{0}^{\infty}dp_{1}\sinh(\pi
p_{1})\exp(-p_{1}\vert x_{0}-x_{0}^{\prime}\vert) \sum_{k\neq
0}\vert K_{ip_{1}}(\omega_{k}\exp(x_{1}))\vert^{2}
\end{array}\end{equation}
Let $\epsilon_{1}$ be the lowest non-zero eigenvalue. The finite
sum on the rhs of eq.(38) is decreasing as
$\exp(-2\sqrt{m^{2}+\epsilon_{1}}\exp(x_{1}))$. For this reason we
can begin the sum starting from large eigenvalues. For large
eigenvalues ($\epsilon_{k}\geq n$ with $n$ sufficiently large) we
can apply the Weyl approximation for the eigenvalues
 distribution \cite{taylor} with the conclusion
\begin{equation}\begin{array}{l}
\vert{\cal G}^{m}(x_{0},x_{1},z;x_{0}^{\prime},x_{1},z)-
g_{0}^{m}(x_{0},x_{1};x_{0}^{\prime},x_{1}) \vert \cr \leq
C^{2}\frac{4}{\pi^{2}}\int_{0}^{\infty}dp_{1}\sinh(\pi
p_{1})\exp(-p_{1}\vert x_{0}-x_{0}^{\prime}\vert) \sum_{\delta
\leq \epsilon_{k}\leq n}\vert
K_{ip_{1}}(\omega_{k}\exp(x_{1}))\vert^{2}\cr+
R\int_{0}^{\infty}dp_{1}\sinh(\pi p_{1})\exp(-p_{1}\vert
x_{0}-x_{0}^{\prime}\vert) \int_{\vert {\bf k}\vert\geq
\sqrt{n}}d{\bf k}\vert K_{ip_{1}}(\sqrt{\vert{\bf
k}\vert^{2}+m^{2}}\exp(x_{1}))\vert^{2}
\end{array}\end{equation}
The finite sum as well as the integral on the rhs of eq.(39) are
decaying exponentially in $x_{1}$ with the rate
$\sqrt{\epsilon_{1}+m^{2}}$ whereas for $g_{0}^{m}$ we can get the
estimate
\begin{displaymath}
\vert g_{0}^{m}(x_{0},x_{1};x_{0}^{\prime},x_{1})\vert \leq
K\exp\Big(-m \vert x_{0}-x_{0}^{\prime}\vert-m\exp(x_{1})\Big)
\end{displaymath}
for non-negative $x_{1}$. Hence, for any $m\geq 0$ the rhs of
eq.(13) is decreasing to zero faster than both ${\cal G}^{m}$ and
$g_{0}^{m}$.
\section{Green functions  on a product manifold}
In secs.2 and 3 we have approximated a manifold with a bifurcate
Killing horizon by a product manifold and discussed the Green
functions in such an approximation. In this section we consider
Green functions on product manifolds in a formulation based on our
earlier paper \cite{haba-cqg}. Here, we emphasize some methods
which have applications to the black brane solutions to be
discussed in the next section.

Let us consider  a manifold in the form of a product ${\cal N}={\cal
K}\times {\cal M}$ where ${\cal K}$ has $D-d+1$ dimensions and
${\cal M}$ is a $d$ dimensional manifold. The metric on ${\cal N}$
takes the form
\begin{equation}
ds^{2}=\sigma_{AB}dX^{A}dX^{B}=g_{ab}(w)dw^{a}dw^{b}+h_{jk}(z)dz^{j}dz^{k}
\end{equation}
where the coordinates  on ${\cal N}$ are denoted by the capital
$X=(w,z)$, the ones on ${\cal K}$ by $w$ and the coordinates on
${\cal M}$ are denoted by $z$. A solution of eq.(4) can be
expressed by the fundamental solution
 of the diffusion equation
\begin{equation}
\partial_{\tau}P_{\tau}^{N}=\frac{1}{2}\triangle_{N}P_{\tau}^{N}
\end{equation}
with the initial condition
$P_{0}(X,X^{\prime})=\sigma^{-\frac{1}{2}}\delta(X-X^{\prime})$.
Then
\begin{equation}
{\cal
G}_{N}^{m}=\frac{1}{2}\int_{0}^{\infty}d\tau\exp(-\frac{1}{2}m^{2}\tau)
P_{\tau}^{N}
\end{equation}We may write eq.(4) in the form (here $
h_{M}=\det(h_{jk})$ and $g=\det(g_{ab})$)
\begin{equation}
(-\triangle_{N}+m^{2} ){\cal
G}_{N}^{m}=(-\triangle_{M}+m^{2}-\triangle_{K} ){\cal
G}_{N}^{m}=h_{M}^{-\frac{1}{2}}g^{-\frac{1}{2}}\delta( X-
X^{\prime})
\end{equation}
From eq.(43) we have a  simple formula ( in the sense of a product
of semigroups)
\begin{equation}
P_{\tau}^{N}=P_{\tau}^{K}P_{\tau}^{M} \end{equation} where the
upper index of the heat kernel denotes the manifold of its
definition.

 Hence
\begin{equation}
{\cal
G}_{N}^{m}(X,X^{\prime})=\frac{1}{2}\int_{0}^{\infty}d\tau\exp(-\frac{1}{2}m^{2}\tau)
P_{\tau}^{K}(w,w^{\prime})P_{\tau}^{M}(z,z^{\prime})
\end{equation}

We expand the Green function (distinguishing the zero mode) in
eigenfunctions $u_{k}$ (12) of the Laplace-Beltrami operator
$\triangle_{M}$
\begin{equation}\begin{array}{l}
{\cal G}_{N}^{m}(X,X^{\prime})-g^{m}_{0}(w,w^{\prime})=
\sum_{k\neq
0}g_{k}^{m}(w,w^{\prime})\overline{u}_{k}(z)u_{k}(z^{\prime})\end{array}
\end{equation}
$g_{k}^{m}$ is a solution of the equation
 \begin{equation}
 {\cal A}_{k}g_{k}^{m}(w,w^{\prime})=
\Big(\sqrt{g}\omega_{k}^{2}-
\partial_{a}g^{ab}\sqrt{g}\partial_{b}\Big)g^{m}_{k}=\delta(w-w^{\prime})
 \end{equation}
where
   $\omega_{k}$  is defined in eq.(9).
The zero mode $g_{0}^{m}$ is the solution of the equation
\begin{equation}
\Big(-
\partial_{a}g^{ab}\sqrt{g}\partial_{b}+m^{2}\sqrt{g}\Big)g^{m}_{0}
=\delta(w-w^{\prime})
\end{equation}
As in sec.3 we ask the question   whether ${\cal G}^{m}_{N}$ can
be approximated by $g_{0}^{m}$. As a first rough approximation for
large distances we apply the WKB reresentation expressing the
Green function $g_{k}$ (47) in the form
\begin{equation}
g_{k}^{m}(w,w^{\prime})=\exp(-\omega_{k}W(w,w^{\prime}))
\end{equation}
Assuming that $W$ is growing uniformly in each direction we obtain
 in the leading order for large distances
 the equation
\begin{equation}
1=g^{ab}\partial_{a}W\partial_{b}W
\end{equation}
 for $W$. Eq.(50) is an equation
for the geodesic distance $\sigma_{K}$ on the manifold ${\cal K}$
with the metric $g_{ab}$ \cite{dewitt}. Hence, the geodesic
distance $W(w,w^{\prime})=\sigma_{K}(w,w^{\prime})$ is the
solution of eq.(50) which is symmetric under the exchange of the
points and satisfies the boundary condition $W(w,w)=0$. We insert
the approximate solutions $g_{k}^{m}$ (49)  into the sum (46).
Then, we can express the sum by the heat kernel of
$\sqrt{-\triangle_{M}+m^{2}}$ or the heat kernel of
$-\triangle_{M}+ m^{2}$

\begin{equation}\begin{array}{l}
{\cal G}_{N}^{m}(X,X^{\prime}) -g_{0}(w,w^{\prime})\simeq
\sum_{k}\exp(-\omega_{k}W(w,w^{\prime}))\overline{u}_{k}(z)u_{k}(z^{\prime})
\cr
=\exp(-\sqrt{-\triangle_{M}+m^{2}}W(w,w^{\prime}))(z,z^{\prime})\cr
= (2\pi)^{-\frac{1}{2}}\int_{-\infty}^{\infty}
d\theta\exp(-\frac{\theta^{2}}{2})\exp\Big(-\frac{W^{2}(w,w^{\prime})}{2\theta^{2}}(-\triangle_{M}+m^{2})\Big)
(z,z^{\prime})\end{array}
\end{equation}
Eq.(51) gives a better approximation than eq.(49) (together with
eq.(50)) of the Green function on ${\cal N}$ because the sum over
eigenvalues $\epsilon_{k}$ has been performed. If the heat kernel on
${\cal M}$ is known (to be discussed in the next section) then from
eq.(51) we can obtain an approximation for ${\cal G}^{m}_{N}$.

\section{ Green functions for approximate geometries of black holes and black branes}
In sec.3 we have derived an approximation for the Riemannian
metric near the bifurcate Killing horizon. We could see that at
the  horizon $ {\cal N}\simeq {\cal R}_{2}\times {\cal M}$ . In
the case of the Schwarzschild black hole in four dimensions we
have
 $ {\cal N}\simeq {\cal R}_{2}\times S^{a}_{2}$. We are interested in this section also in the black
  brane solutions  in 10 dimensional supergravity which have the near horizon geometry $ {\cal
N}\simeq AdS_{5}\times S_{5}^{a}$ and the black brane solutions in
11 dimensional supergravity with the near horizon geometry $ {\cal
N}\simeq AdS_{7}\times S_{4}^{a}$ or $ {\cal N}\simeq
AdS_{4}\times S_{7}^{a}$ \cite{duff}. These manifolds can be
considered as solutions to the string theory  compactification
problem \cite{becker}. We can see that for a description of the
black hole (black brane) Green functions the formulas for the heat
kernel on the Rindler space and AdS space will be useful. We shall
set the radius of the black hole (brane) $a=1$ in this section.
The radius can be inserted in our formulas by a restoration of
proper dimensionality of the numbers entering these formulas (as
will be indicated further on). For the Rindler space the heat
kernel equation reads
\begin{equation}
\partial_{\tau}P^{R}_{\tau}
=\frac{1}{2}\triangle_{R}P^{R}_{\tau}=\frac{1}{2}(y^{-2}\partial_{0}^{2}+y^{-1}\partial_{y}y\partial_{y})P^{R}_{\tau}
\end{equation}
If we take the Fourier transform in $x_{0}$ in eq.(52) then we
obtain an equation for the Bessel function $I_{\vert p_{0}\vert}$
\cite{grad}. Hence, we  can see that the solution of the heat
kernel for ${\cal R}_{2}$ with the initial condition
$(yy^{\prime})^{-\frac{1}{2}}\delta$ can be expressed in the form
\begin{equation}
P_{\tau}^{R}(x,y;x^{\prime},y^{\prime})=\frac{1}{\pi\tau}(2\pi)^{-\frac{1}{2}}\int
dp_{0} \exp(ip_{0}(x_{0}-x_{0}^{\prime}))I_{\vert
p_{0}\vert}(\tau^{-1}yy^{\prime})\exp(-\frac{1}{2\tau}(y^{2}+y^{\prime
2}))\end{equation}  Eq.(52) for the heat kernel $P_{\tau}^{R}$ on
${\cal R}_{2}$ coincides with the heat equation on the plane but
expressed in cylindrical coordinates,i.e., if
\begin{displaymath}w_{0}=y\cos x_{0},w_{1}=y\sin x_{0}
\end{displaymath}
then \begin{displaymath}
\triangle_{R}=\frac{\partial^{2}}{\partial
w_{0}^{2}}+\frac{\partial^{2}}{\partial w_{1}^{2}}
\end{displaymath}
As a consequence the heat kernel $P_{\tau}^{2\pi}$ with periodic
boundary conditions imposed on $x_{0}$( with the period $2\pi$ )
must coincide with the heat kernel on the plane $R^{2}$ (see also
\cite{linet}). Hence,
\begin{equation}
P_{\tau}^{2\pi}=(2\pi\tau)^{-1}\exp\Big(-\frac{1}{2\tau}\vert w-w^{\prime}\vert^{2}\Big)
\end{equation}
where
\begin{equation}
\vert w-w^{\prime}\vert^{2}=(w_{0}-w_{0}^{\prime})^{2}
+(w_{1}-w_{1}^{\prime})^{2}
\end{equation}
First, we apply the methods of sec.4  to the case discussed already
in another way in sec.3.  The approximate near horizon geometry of
the Schwarzschild black hole in four dimensions is ${\cal N}={\cal
R}_{2}\times S_{2}$(in $D+1$ dimensions this is ${\cal N}={\cal
R}_{2}\times S_{D-1}$). We may apply  eq.(45) in order to express
the Green function by the heat kernels. For this purpose the
eigenfunction expansion on $S_{2}$ is useful
\begin{equation}
P_{\tau}^{S_{2}}(\sigma)=\frac{1}{4\pi}\sum_{l=0}^{\infty}(2l+1)P_{l}(\cos
\sigma)\exp(-\frac{\tau}{2}l(l+1))
\end{equation}
where $P_{l}$ is the Legendre polynomial and the geodesic distance
$\sigma$ is defined in eq.(33). Applying the representation (34)
of the Legendre polynomials  we can sum up the series (56) and
express it by an integral
\begin{equation}\begin{array}{l}
P_{\tau}^{S_{2}}(\sigma)=\frac{1}{4\pi^{2} } (2\pi
\tau)^{-\frac{1}{2}}\int_{0}^{\pi}d\phi\int_{-\infty}^{\infty}
du\exp(-\frac{u^{2}}{2\tau}) \cr (1+\Omega )(1-\Omega)^{-2}
\end{array}\end{equation}
where
\begin{equation}
\Omega=\exp(iu-\frac{\tau}{2})(\cos\sigma+i\sin\sigma\cos\phi)
\end{equation}
If we expand $(1-\Omega)^{-2}$ in $\Omega$ then we obtain the
expansion (56) of the heat kernel.

Applying eqs.(45),(54) and (57) we can represent the scalar Green
function on the four-dimensional black hole of temperature
$\beta=2\pi$ ( this is the conventional Hawking temperature in
dimensionless units; the $x_{0}$ coordinate is made an angular
variable in order to make the singular conical manifold regular
\cite{hawking-gibbons}\cite{becker}\cite{cone}\cite{kirsten})
\begin{equation}
\begin{array}{l}
{\cal
G}^{m}_{2\pi}(w,\theta,\phi;w^{\prime},\theta^{\prime},\phi^{\prime})-g^{m}_{0}(w,w^{\prime})=\frac{1}{2\pi^{2}
}\int_{0}^{\pi}d\phi\int du\int_{0}^{\infty} d\tau (2\pi
\tau)^{-\frac{3}{2}}\exp(-\frac{1}{2}m^{2}\tau)\cr\exp\Big(-\frac{1}{2\tau}(\vert
w-w^{\prime}\vert^{2}+u^{2})\Big) \Omega(1-\Omega)^{-2}
\end{array}
\end{equation}
where
\begin{displaymath}
(-\frac{\partial^{2}}{\partial
w_{0}^{2}}-\frac{\partial^{2}}{\partial
w_{1}^{2}}+m^{2})g_{0}^{m}=\delta(w-w^{\prime})
\end{displaymath}
We obtain an exponential decay of the rhs of eq.(59) if we make the
approximation $(1-\Omega)^{-2}\simeq 1$ and estimate the correction
to this approximation. In this way we reach the approximation (37)
of sec.3 but now at the Hawking temperature. For zero temperature or
a temperature different from the Hawking temperature it would be
difficult to obtain a useful approximation for the Green function
from eq.(45) because the formula (53) for the heat kernel on the
Rindler space is rather implicit.

   We can obtain simple  formulas for the
Schwarzschild black hole in an odd dimension $d+2$. For an odd $d$
the formula for the heat kernel  on $S_{d}$ reads \cite{campo}
\begin{equation}\begin{array}{l}
P_{\tau}^{S_{d}}=(\frac{1}{2\pi})^{\frac{d-1}{2}}\exp(\frac{(d-1)^{2}\tau}{8})\Big(\frac{\partial}{\partial
\cos\sigma}\Big)^{\frac{d-1}{2}}(2\pi\tau)^{-\frac{1}{2}}\sum_{n}
\exp(-\frac{1}{2\tau}(\sigma-2\pi n)^{2}) \cr
=(\frac{1}{2\pi})^{\frac{d-1}{2}}\exp(\frac{(d-1)^{2}\tau}{8})\Big(\frac{\partial}{\partial
\cos\sigma}\Big)^{\frac{d-1}{2}}(2\pi)^{-1}\sum_{2 n \geq d-1}
\exp(-\frac{\tau n^{2}}{2})\cos(n\sigma)
\end{array}\end{equation}
(the derivative over $\cos \sigma$ annihilates all $\cos n\sigma$
with $2n<d-1$).
 Performing the $\tau$ integral (45)(of the heat kernels (54) and (60)) we obtain the Green function of the black hole in $d+1$
 dimensions at temperature $\beta=2\pi$
\begin{equation}
\begin{array}{l}
{\cal G}_{2\pi}^{m}(w,w^{\prime};\sigma)-g^{m}_{0}(w,w^{\prime})
=\frac{2}{\pi}(\frac{1}{2\pi})^{\frac{d-1}{2}}\Big(\frac{\partial}{\partial
\cos\sigma}\Big)^{\frac{d-1}{2}} \cr\sum_{2 n> d-1} K_{0}\Big(
(n^{2}-\frac{1}{4}(d-1)^{2}+m^{2})^{\frac{1}{2}}\vert
w-w^{\prime}\vert\Big)\cos(n\sigma)
\end{array}\end{equation}
We can use the representation (27) of the Bessel function
\begin{displaymath}
K_{0}(u)=\int_{0}^{\infty}dt \exp(-u\cosh(t))
\end{displaymath}
in order to sum the series (61) for large $n\geq\Lambda$. Then, we
can use the approximation
$(n^{2}-\frac{1}{4}(d-1)^{2}+m^{2})^{\frac{1}{2}}\simeq\vert
n\vert$. In such a case
\begin{equation}
\begin{array}{l}
{\cal G}^{m}_{2\pi}(w,w^{\prime};\sigma)-g^{m}_{0}(w,w^{\prime})
\simeq\frac{2}{\pi}(\frac{1}{2\pi})^{\frac{d-1}{2}}\Big(\frac{\partial}{\partial
\cos\sigma}\Big)^{\frac{d-1}{2}} \cr\Big(\sum_{2 n>
d-1}^{n<\Lambda} \cos(n\sigma)K_{0}\Big(
(n^{2}-\frac{1}{4}(d-1)^{2}+m^{2})^{\frac{1}{2}}\vert
w-w^{\prime}\vert\Big)\cos(n\sigma) \cr +\int_{0}^{\infty} dt\Big(
\exp(i\Lambda\sigma-\Lambda \vert w-w^{\prime}\vert\cosh(t))
\Big(1-\exp(i\sigma- \vert w-w^{\prime}\vert\cosh(t))\Big)^{-1}
\cr +\exp(-i\Lambda\sigma-\Lambda \vert w-w^{\prime}\vert\cosh(t))
\Big(1-\exp(-i\sigma-\vert
w-w^{\prime}\vert\cosh(t))\Big)^{-1}\Big)
\end{array}\end{equation}The finite
number of terms on the rhs of eq.(62) is decreasing exponentially
because $K_{0}(u)$ is decreasing exponentially for large $u$.
Then, the integral over $t$ is decreasing exponentially as in the
definition of the Bessel function (27). In order to prove this we
make the approximation \begin{displaymath} \Big(1-\exp(i\sigma-
\vert w-w^{\prime}\vert\cosh(t))\Big)^{-1} \simeq 1
\end{displaymath}
and subsequently estimate the correction to this approximation.
 We
can conclude that  the rhs of eq.(62) is decreasing as
$\exp(-\sqrt{m^{2}+1}\vert w-w^{\prime}\vert) $

(as $\exp(-\sqrt{m^{2}+a^{-2}}\vert w-w^{\prime}\vert) $ after an
insertion of the radius $a$ of the sphere).

We obtain explicit formulas for scalar Green functions on black
brane solutions of supergravity in ten dimensions which near the
horizon have the $AdS_{5}\times S_{5}$ geometry
\cite{brane}\cite{maldacena}\cite{duff}. The heat kernel on the
$2k+3$ dimensional hyperbolic space ${\cal H}_{2k+3}$(Euclidean
$AdS_{2k+3}$ ) is \cite{grig}\cite{haba-jpa}
 \begin{equation}\begin{array}{l}
p_{\tau}^{(k+1)}(\sigma_{H})
=-(-2\pi)^{-k-1}\exp(-\frac{(k+1)^{2}}{2}\tau+\frac{1}{2}\tau)
\cr\Big(\frac{d}{d\cosh\sigma_{H}}\Big)^{k}p_{\tau}^{(1)}(\sigma_{H})
\end{array}\end{equation} with \begin{equation}
p_{\tau}^{(1)}(\sigma_{H})=(2\pi
\tau)^{-\frac{3}{2}}\sigma_{H}(\sinh\sigma_{H})^{-1}\exp(-\frac{\tau}{2}-\frac{\sigma_{H}^{2}}{2\tau})
\end{equation}
It is a function of the Riemannian distance $\sigma_{H}$. In the
Poincare coordinates $(y,{\bf x})$
\begin{displaymath}\cosh
\sigma_{H}=1+(2yy^{\prime})^{-1}(({\bf x}-{\bf
x}^{\prime})^{2}+(y-y^{\prime})^{2})\end{displaymath} The integral
(45) over $\tau$ can be calculated using eqs.(60) and (63). Then,
 for ${\cal H}_{5}\times S_{5}$ we obtain
\begin{equation}
\begin{array}{l}{\cal G}^{m}_{5+5}(\sigma_{H},\sigma)-
g_{05}^{m}(\sigma_{H})=\frac{1}{4\pi^{3}}(2\pi)^{-\frac{3}{2}}\Big(\frac{\partial}{\partial
\cos\sigma}\Big)^{2}\frac{\partial}{\partial \cosh \sigma_{H}}\cr
(\sinh\sigma_{H})^{-1}\sum_{ n>2}(
n^{2}+m^{2})^{\frac{1}{2}}\exp\Big(- (
n^{2}+m^{2})^{\frac{1}{2}}\sigma_{H}\Big) \cos(n\sigma)
\end{array}\end{equation}
where $g_{05}^{m}$ is the Green function on the hyperbolic space
${\cal H}_{5}$. In the Poincare coordinates $g_{05}^{m}$ is the
solution of the equation
\begin{displaymath}
(-y^{2}\partial_{y}^{2}+3y\partial_{y}-y^{2}\triangle_{{\bf x}}
+m^{2})g_{05}^{m}=y^{5}\delta \end{displaymath} We have
\begin{equation}
g_{05}^{m}(\sigma_{H})=-\frac{1}{2\pi}\frac{\partial}{\partial \cosh
\sigma_{H}} (\sinh\sigma_{H})^{-1}
\exp\Big(-\sqrt{4+m^{2}}\sigma_{H}\Big)
\end{equation}
If the sphere has radius $a$ and the hypersphere the radius $b$ then
the exponential factor in eq.(66) reads
\begin{displaymath}
\exp\Big(-b\sqrt{4a^{-2}+m^{2}}\sigma_{H}\Big)
\end{displaymath}
 We can  sum on the rhs of
eq.(65) a finite number of terms and subsequently approximate the
remaining series replacing the square root in the argument of the
exponential by $\vert n\vert$ for large $n$ (the procedure which we
applied at eq.(62)).
 If the mass $m=0$ then the sum in eq.(65) can be calculated
exactly with the result
\begin{equation}
\begin{array}{l}{\cal G}^{m}_{5+5}(\sigma_{H},\sigma)-
g_{05}^{m}(\sigma_{H})=\frac{1}{4\pi^{3}}(2\pi)^{-\frac{3}{2}}\Big(\frac{\partial}{\partial
\cos\sigma}\Big)^{2}\Big(\frac{\partial}{\partial \cosh
\sigma_{H}}\Big)^{2}\cr
\exp(-3\sigma_{H})(\cos(3\sigma)-\exp(-\sigma_{H})\cos(2\sigma))\Big(1+\exp(-2\sigma_{H})-2
\exp(-\sigma_{H})\cos(\sigma)\Big)^{-1}\end{array}\end{equation}

In eleven dimensions the counterpart of the ten dimensional black
brane has the near horizon geometry $AdS_{7}\times S_{4}$. The
heat kernel on $S_{4}$ can be expressed in the form \cite{campo}
\begin{equation}
P_{\tau}^{S_{4}}(\sigma)=\frac{1}{8\pi^{2}}\frac{d}{d\cos\sigma}\sum_{l=0}^{\infty}(2l+1)P_{l}(\cos\sigma)
\exp(-\frac{\tau}{2}l(l+1))
\end{equation}
where $\sigma$ is the geodesic distance on $S_{4}$. Hence, from
eq.(45)
\begin{equation}
\begin{array}{l}{\cal G}_{7+4}^{m}(\sigma_{H},\sigma)-
g_{07}^{m}(\sigma_{H})=\frac{1}{8\pi^{2}}(2\pi)^{-3}(2\pi)^{-\frac{3}{2}}\frac{\partial}{\partial
\cos\sigma}\Big(\frac{\partial}{\partial \cosh
\sigma_{H}}\Big)^{2} (\sinh\sigma_{H})^{-1}\cr\sum_{l>0}\Big(
l(l+1)+m^{2}+9\Big)^{\frac{1}{2}}\exp\Big(-\Big(
l(l+1)+m^{2}+9\Big)^{\frac{1}{2}}\sigma_{H}\Big)P_{l}(\sigma)
\end{array}\end{equation}
where $g_{07}^{m}$ is the Green function on the seven dimensional
hyperbolic space ${\cal H}_{7}$. It is the solution of the
equation (in Poincare coordinates)
\begin{equation}
(-y^{2}\partial_{y}^{2}+5y\partial_{y}-y^{2}\triangle_{{\bf x}}
+m^{2})g_{07}^{m}=y^{7}\delta \end{equation} The solution of
eq.(70) reads
\begin{equation}
g_{07}^{m}(\sigma_{H})=(2\pi)^{-2}\Big(\frac{\partial}{\partial
\cosh \sigma_{H}}\Big)^{2} (\sinh\sigma_{H})^{-1}
\exp(-\sqrt{9+m^{2}}\sigma_{H}) \end{equation}
 Again we can
obtain good approximation
 for the Green function summing a finite number of terms on the
 rhs of eq.(69)and subsequently calculating the infinite sum with the
 approximation $\Big(
l(l+1)+m^{2}+9\Big)^{\frac{1}{2}}\simeq l$.

Finally, let us consider  the black brane solution of supergravity
in eleven dimensions with the near horizon geometry $AdS_{4}\times
S_{7}$. The Green functions have a more involved representation in
this case because the heat kernel on the even dimensional
hyperbolic space cannot be expressed by elementary functions. We
have
\begin{equation}
p_{\tau}=-\sqrt{2}(2\pi)^{-1}\frac{d}{d\cosh
\sigma_{H}}\exp(-\frac{9}{8}\tau)(2\pi\tau)^{-\frac{3}{2}}
\int_{\sigma_{H}}^{\infty}(\cosh r-\cosh
\sigma_{H})^{-\frac{1}{2}} r\exp(-\frac{r^{2}}{2\tau})dr
\end{equation}
Hence, from eqs.(45)and (65) we obtain
\begin{equation}
\begin{array}{l}
{\cal
G}_{4+7}^{m}(\sigma_{H},\sigma)-g_{04}^{m}(\sigma_{H})=-\sqrt{2}(2\pi)^{-4}\frac{d}{d
\cosh \sigma_{H}}\Big(\frac{d}{d{\cos
\sigma}}\Big)^{3}\sum_{n>3}\sqrt{-\frac{27}{4}+m^{2}+n^{2}} \cr
\int_{\sigma_{H}}^{\infty}dr(\cosh r-\cosh
\sigma_{H})^{-\frac{1}{2}}\exp\Big(-r\sqrt{-\frac{27}{4}+m^{2}+n^{2}}\Big)
\end{array}\end{equation}
where \begin{equation}\begin{array}{l}
g^{m}_{04}(\sigma_{H})=-2\sqrt{2}(2\pi)^{-1}\frac{d}{d \cosh
\sigma_{H}} \int_{\sigma_{H}}^{\infty}(\cosh r-\cosh
\sigma_{H})^{-\frac{1}{2}} \exp(-r\sqrt{\frac{9}{4}+m^{2}})dr
\end{array}\end{equation}
The integrals over $r $ in eqs.(72)-(74) can be expressed by the
Legendre functions $Q_{\alpha}$ using the integral representation
\cite{grad}\begin{displaymath} Q_{\alpha}(\cosh
\sigma)=\int_{\sigma}^{\infty}dr(2\cosh r-2\cosh
\sigma)^{-\frac{1}{2}}\exp(-\alpha r-\frac{1}{2}r)
\end{displaymath}

\section{Discussion} It is known that the Riemannian geometry of the spherically symmetric
black hole  manifold of $D+1$ dimensions near the horizon can be
approximated by ${\cal N}={\cal R}_{2}\times S_{D-1}$. The black
brane solutions of supergravity and string theory near the horizon
can be approximated by a product ${\cal N}=AdS_{p}\times S_{q}$.
Let us denote these manifolds with a horizon by $\tilde{{\cal
N}}$. A question could be raised to what extent the product
manifold ${\cal N}={\cal K}\times {\cal M}$  is a good
approximation to $\tilde{{\cal N}}$ in the sense that $\vert {\cal
G}_{N}(X,X^{\prime})-\tilde{{\cal G}}_{N}(X,X^{\prime})\vert$ is
small in a certain range of $X$ and $X^{\prime}$. Such problems
have been studied for the heat kernels in relation to the
parametrix method for diffusion equations \cite{pogo}\cite{mol}.
By these methods we could estimate $\vert
P_{\tau}^{N}-\tilde{P}_{\tau}^{N}\vert$ and subsequently integrate
the estimate over $\tau$ (in the sense of eq.(42)). However,
explicit estimates  would be difficult and are beyond the scope of
this work. We restricted ourselves in secs.3 and 5 to an answer
(by different methods) to a simpler problem: whether the Green
function on a product manifold can be approximated by its zero
mode. We have obtained explicit formulas for the correction to the
zero mode contribution. The zero mode is the Green function on the
non-compact part. We have shown that the scalar propagator on the
product manifold can indeed be approximated for large distances by
the one on the non-compact part  of ${\cal N}$. The contribution
of the compact part is decreasing exponentially as a function of
the distance. This is in fact an expression of the
compactification in the Kaluza-Klein setting. In the case of the
Schwarzschild black hole we could call such a  phenomenon a
dimensional reduction.
 The compact degrees of freedom become irrelevant for the
scalar  quantum field theory defined on the black hole background.
The  ${\cal N}\simeq AdS_{p}\times S_{q}$ approximation is usually
discussed \cite{witten}\cite{maldacena} in  relation to an
approximation by a conformal field theory on the boundary of
$AdS_{p}$. The form of the propagator on the product of the
non-compact and compact manifolds could be useful for a complete
reconstruction of the quantum field theory on ${\cal N}$ from the
one on the boundary of $AdS_{p}$. Some consequences of the near
horizon dimensional reduction in the case of the Schwarzschild
black hole have been discussed by Padmanabhan \cite{padma}.

\end{document}